\RequirePackage{fix-cm}

\documentclass[sigconf,screen,nonacm]{acmart}        
\usepackage{graphicx}
\usepackage{natbib}
\usepackage{float}

\begin{document}

\title{The Uncanny Valley in medical simulation-based training: a visual summary}

\author{Eleni Grigoriou}
\affiliation{%
  \institution{University of Crete Greece, ORamaVR Switzerland}
  \city{}
  \country{}
}

\author{Manos Kamarianakis}
\affiliation{%
  \institution{FORTH - ICS Greece, University of Crete Greece, ORamaVR Switzerland}
  \city{}
  \country{}
}

\author{George Papagiannakis}

\affiliation{%
  \institution{FORTH - ICS Greece, University of Crete Greece, ORamaVR Switzerland}
  \city{}
  \country{}
}

\date{ }

\maketitle

\section{Introduction}
The purpose of this review article is to provide a \textbf{bibliographical} as well as \textbf{evidence-based visual guide} regarding the effect of \textbf{``Uncanny Valley'' (UV)} and how it profoundly influences \textbf{medical virtual reality} simulation-based \textbf{training}. The phenomenon, where increasingly realistic virtual humans elicit discomfort due to subtle imperfections, is crucial to understand and address in the context of medical training, where realism and immersion are key to effective learning.

Our research team, consisting of experts in computer graphics, virtual reality, and medical education, brings a diverse and multidisciplinary perspective to this subject. Our collective experience spans developing advanced computer graphics systems, VR character simulation, and innovative educational technologies. We have collaborated across institutions and industries to push the boundaries of VR applications in medical training.

\section{Background and definitions}

\subsection{Anthropomorphism as a driving force }

Humans use the technique of anthropomorphism to attribute humanlike traits, emotions or intentions to inanimate objects and animals. This allows us to rationalize a character’s actions and accept them as a social companion, despite them being a nonhuman entity. For example, in the same way that we may attribute human-like characteristics and a personality to one’s pet dog or cat, we can assign cognitive (thoughts) and emotional states to a nonhuman-like object. As \cite{Mori1970} predicted, empirical research has identified that anthropomorphic character designs that lack a nonhuman appearance, but that portray human-like traits, are regarded more positively than realistic, human-like, believable characters \cite{Yu2011,Jung2011,Tinwell2014}.

\subsection{Photorealism and physically based rendering for architectural scenes}

The purpose of computer graphics has always been the generation of computer imagery based on mathematical models. Naturally, this process has been extended towards physically based rendering \cite{Phar-2017} to reach such realism that is indistinguishable from photographs (Photorealism) \cite{IKEA2014}. Nowadays we have established the complete mathematical framework to physically simulate light interaction in a synthetic scene in order to depict photorealistic computer graphics (CG) images \cite{Phar-2017}, albeit we lack the computational resources to calculate them in real-time (interactively). However, this is not the case for non-interactive exterior or interior architectural static (light or camera constrained) scenes since photorealism has been achieved in these cases, as figures illustrated below. In the case though of interactive VR walkthroughs, the computational power still is elusive and further approximations are needed \cite{Tato2012}.

\begin{figure}
    \centering
    \includegraphics[width=0.85\linewidth]{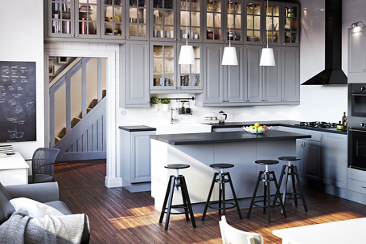}
    \caption{An example of non-realtime, photorealistic, physically based rendering (still image) of an interior architectural scene from \cite{IKEA2014} where more than 75\% of their catalogue is now computer-graphics based, as opposed to traditional photography}
    \label{fig:fig1}
\end{figure}

\begin{figure}
    \centering
    \includegraphics[width=0.85\linewidth]{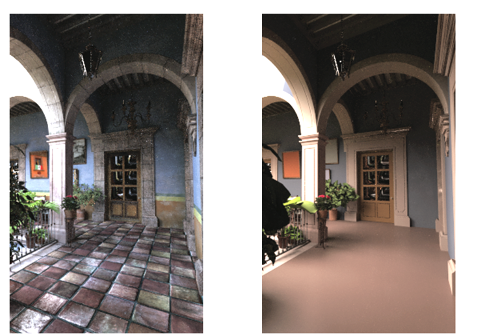}
    \caption{Another example of off-line, synthetic, photorealistic exterior architectural scene from \cite{Phar-2017} as a still image employing global illumination methods and textures (left) as opposed to only colour-based physical based materials (right)}
    \label{fig:fig2}
\end{figure}

\begin{figure}
    \centering
    \includegraphics[width=0.45\linewidth]{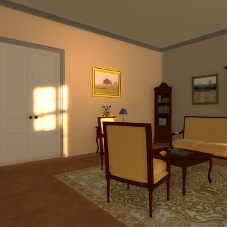}
    \includegraphics[width=0.45\linewidth]{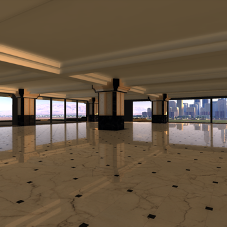}
    \caption{Real-time global illumination approximations employed in VR architectural walkthroughs \cite{Tato2012}. These are still images taken from an interactive simulation where the user is controlling the virtual camera while freely moving in the virtual environment in real-time}
    \label{fig:fig3}
\end{figure}

\subsection{Virtual Reality for training}

There are currently three different working definitions of virtual reality that we endorse: 
\begin{itemize}
\item VR (1): An experience that requires a headset to completely replace a user’s surrounding view with a simulated, immersive, and interactive virtual environment. \cite{Fink2018}
\item VR (2): The substitution of the interface between a person and the physical environment with an interface to a simulated environment. \cite{Lanier-2017}
\item VR (3): An artificial environment that is immersive enough to convince you that you’re actually inside it. \cite{Rubin2018}
\end{itemize}

Nowadays, a host of virtual simulators for training specialists with difficult jobs, such as astronauts, soldiers, and surgeons have cropped up on the market \cite{Fink2018,Rubin2018,Papagiannakis2018, Papagiannakis2020,Zikas2020}. If these fields have turned to VR, it is for the same reason that flight simulators became invaluable for training pilots. \textbf{``Mistakes are cost and consequence free in VR. And when the risks of on-the-job learning are high, technology that prepares a pilot, or a surgeon, or a soldier for the life-and-death responsibilities of their profession without risk is a huge win''} \cite{Bailenson2018}.

\subsection{Virtual characters and simulated humans}

\cite{Magnenat-Thalmann1987} were amongst the first pioneers to simulate virtual human actors in virtual reality with unprecedented for the time virtual clothing, hair, lip-sync, speech simulation, body and facial expressions. They established the virtual human simulation research field that is active till today. A comprehensive overview of the field is provided in \cite{Thalamnn2012}. Although the question of realism and the work of \cite{Mori1970} on “Uncanny Valley” was well-known, the need for technological advances had superseded then any further deep discussions on that aspect. 

However, nowadays, with the establishment of computer animation movies, 3D games and commercial virtual, augmented and mixed/extended reality simulations, the question of realistic virtual characters is more pertinent than ever: numerous real-time methods are active fields of research for real-time rendering (depiction) of realistic virtual humans \cite{Egges2006,Bailenson2006,Alexander2010}.
“At present, animators may work with a less detailed 3D model of a human-like character at a lower resolution, without seeing full details in skin texture and how light and shadows are cast on the 3D model. The model is then rendered in high definition to achieve full detail in aspects such as folds and wrinkles in a human-like character’s face, skin tone, luminosity and reflection. Jeffrey Katzenburg, the chief executive of the film and animation company DreamWorks, revealed that an experienced animator may take a week to achieve just three seconds of completed animation” \cite{Tinwell2014}.

The automatic (non-manual, designer based) capturing, modelling, rendering and subsequently body deformation and animation of realistic virtual humans has always been a very active field of research. There have been very promising recent results appearing in the bibliography, employing semi-automatic performance capture methods that either require specialized 3D RGB-D cameras \cite{Papaefthymiou2017}, or Light-stage equipment \cite{Seymour2017} with a variety of virtual character enabling technologies on facial animation \cite{Li2017} using machine learning deep generative models \cite{Li2020}, light capture using global illumination \cite{Alexander2010}  and keyframe or physically-based interactive character animation \cite{Jung2011} in realistic VR/AR \cite{Ponder2002}, \cite{Papagiannakis2005b}, \cite{Thalamnn2012}, \cite{Papaefthymiou2015}, \cite{Slater2016}, \cite{Papaefthymiou2017b}. Holographic video virtual humans \cite{Collet2015} are a totally different category (most realistic depictions of virtual humans up to date), but they refer to video-based representations and not real-time interactive virtual characters. \textbf{A comprehensive, integrated solution to employ realistic virtual characters in medical training though has never so far crossed the research field into production-ready medical simulation-based training.}

\begin{figure}
    \centering
    \includegraphics[width=0.35\linewidth]{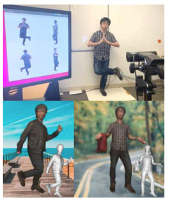}
    \includegraphics[width=0.62\linewidth]{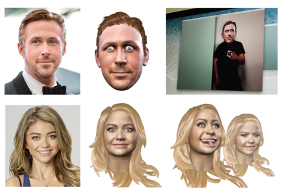}
    \caption{Performance capture and modelling of a realistic virtual human based on video and using deep learning methods in \cite{Li2020} (left) and out of photographs by \cite{Li2017}  (right).}
    \label{fig:fig4}
\end{figure}

\begin{figure}
    \centering
    \includegraphics[width=0.95\linewidth]{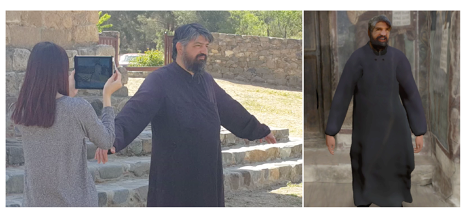}
    \caption{One of our methods for real-time modelling and simulation of virtual humans in VR \cite{Papaefthymiou2017}}
    \label{fig:fig5}
\end{figure}

\begin{figure}
    \centering
    \includegraphics[width=0.95\linewidth]{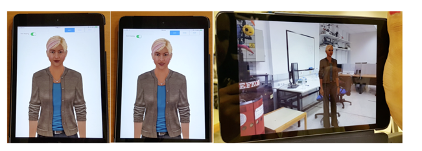}
    \caption{Another of our algorithms for real-time simulation of realistic virtual humans in AR \cite{Papaefthymiou2015}}
    \label{fig:fig6}
\end{figure}

\begin{figure}
    \centering
    \includegraphics[width=0.95\linewidth]{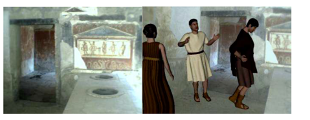}\\
    \includegraphics[width=0.55\linewidth]{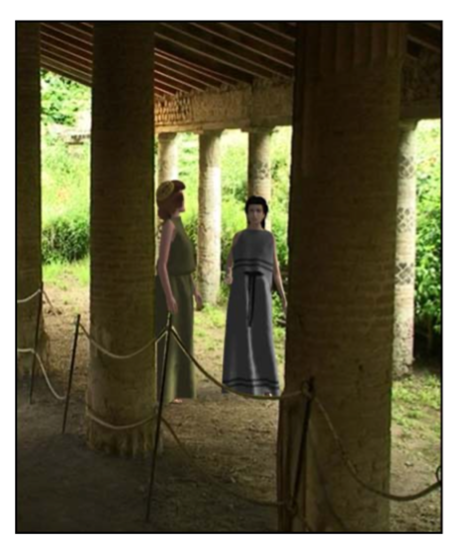}
    \caption{Early examples of several of our algorithms and techniques for realistic virtual human simulation in AR \cite{Papagiannakis2005}}
    \label{fig:fig7}
\end{figure}

\subsection{Presence and the place / plausibility illusion}

One of the fundamental characteristics of VR is the sensation of ``being there,'' wherein `there' corresponds to the place that the software application simulates. It is what researchers call psychological presence \cite{Sanchez-Vives2005}, \cite{Slater2016}, \cite{Slater2020}. When this happens, your motor and perceptual systems interact with the virtual world in a manner akin to how they do in the physical world and it triggers your body to respond as though the experience was real \cite{Rubin2018}. Body and hand movement is a key element in VR and directly connects with cognition. Thus, when you experience a VR simulation, you move your body as if you were experiencing an actual event in the real world. What makes VR different from using a computer or a robotic simulator is that you move your body naturally, as opposed to using a mouse and keyboard, or a haptic device. Hence, learners can leverage what psychologists call embodied cognition. Embodied cognition argues that, while of course the mind is located in the brain, there are other organs in the body that influence cognition and that people learn better by doing than by watching \cite{Johnson-Glenberg2018} \cite{Bailenson2008}.

Hence, VR presents three unique characteristics that can drastically impact education and training in a positive way: 1) the sensation of presence, which allows for the collaboration of multiple dispersed participants/agents in the same environment; 2) the embodied affordances of gesture and manipulation in the 3rd dimension and 3) the plausibility illusion, that events that are happening in the virtual environment are real \cite{Yu2011}. These unique attributes are known as the three profound affordances of VR. New generation of mass-produced, standalone mobile VR head mounted displays with hand controllers are now capable of inducing embodiment and agency through meaningful and congruent movements. They are revolutionizing skill transfer from the virtual to the real, in a manner similar to what flight simulators have been doing for pilot training for decades.

\subsection{The Uncanny Valley (UV)}
The ‘Uncanny Valley’ (UV) effect refers to a sense of unease and discomfort when people look at increasingly realistic virtual humans, i.e., "making promises that you can't keep: great polygons but when virtual humans move they appear as if they have neurological damage. The closer to photo-real, the more uncomfortable these characters make the audience because something is never quite right. There's always something just slightly off that we reject. In many cases, it's the spark in a person's eye that simply cannot be recreated” \cite{Tinwell2014}. 

The term ‘Uncanny Valley’ refers to a graph of emotional reaction against the similarity of a robot to human appearance and movement, as shown in the figure below from \cite{Katsyri2015} based on the original graph from . In place of a “toy robot” we can imagine a virtual character in our case.

\begin{figure}
    \centering
    \includegraphics[width=0.75\linewidth]{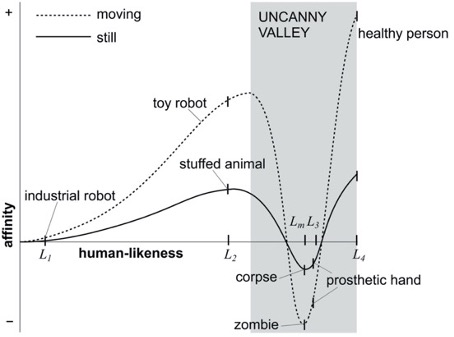}\\
    \includegraphics[width=0.75\linewidth]{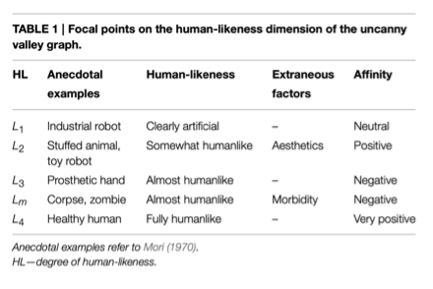}\\
    \includegraphics[width=0.75\linewidth]{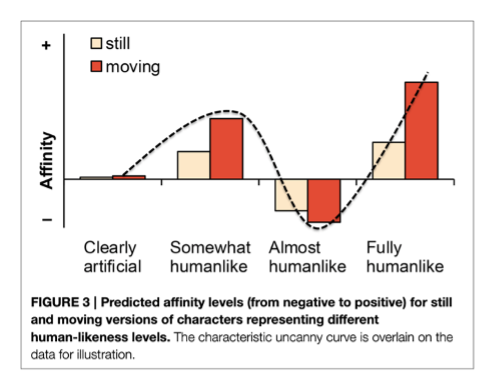}
    \caption{The Uncanny Valley (UV) effect (top left): affinity (empathy, likeness, attractiveness) vs human-likeness from \cite{Katsyri2015} originally adapted from \cite{Mori1970} \cite{Mori2012}}
    \label{fig:fig8}
\end{figure}

The term was coined by the Japanese roboticist Masahiro Mori (1970), although it is often wrongly associated with his later work “The Buddha in the Robot” (1982). As a machine (toy robot or virtual character) acquires greater similarity to a human, it becomes more emotionally appealing to the observer. However, when it becomes disconcertingly close to human there is a very strong drop in believability and comfort, before finally achieving full humanity and eliciting positive reactions once more; this is the Uncanny Valley (UV), depicted in grayscale in the figure above. 

In an essay on the nature of the uncanny, Freud (1919) describes his extreme discomfort feeling of dread and creeping horror, e.g., the mechanical, animatronic doll Olympia from Offenbach's "Tales of Hoffmann". He argues that uncanny reactions occur when something alien is presented in a familiar context or setting. Freud also attempts to identify the word “uncanny” in other languages and their connection notions of unfamiliarity, eeriness, discomfort, disgust or just mysterious response.

“Despite various attempts to simulate human-like facial expression and speech in animation and games (both in prerecorded and real-time footage), there is general agreement that largely such performances can fail to be perceived as real and suspend disbelief for the viewer. As such, the Uncanny Valley is now popular lexicon in the discourse of animation and games, and the phenomenon now occurs as a recognized design issue when creating realistic, human-like characters” \cite{Tinwell2014}

Many horror movie directors have deliberately exploited this phenomenon to heighten an audience's sense of fear and dread. Realistic depiction of lifelike humans is more advanced in film computer-generated imagery (CGI) than within current real time virtual environments. Animators working with CGI characters have described the uncanny as a design limitation which they have had to find creative ways to work around. This is a common secret in the CG world and one of the main reasons that both Pixar™ and Dreamworks™, in all their CG masterpieces up to date, they religiously avoid anthropomorphic, hyper-realistic depictions when human characters are involved in their stories and opt for non-anthropomorphic, more toon-like appearance.

“The word uncanny did describe the chill that I felt up my spine when I realized that there was something not quite right about a CG character presented with a near human-like appearance—to the extent that I was uncomfortable viewing the character’s sinister depiction of a human-like form.” \cite{Tinwell2014}

According to various research examples above, it is clear that it is not the increased realism that elicits the uncanny valley effect, but rather that the increased realism lowers the tolerance for abnormalities and imperfections that realistic VG characters or robots can exhibit.

\section{Research studies on the Uncanny Valley}

There have been several studies so far corroborating the uncanny valley effect \cite{Saygin2012}, \cite{Katsyri2015}, \cite{Wang2017} and the notion that increased anthropomorphism in virtual characters or robots can increase uncanny feelings \cite{Tinwell2014} \cite{Kim2019}.

Research findings from \cite{Katsyri2015} on human-likeness of robots (L1…L4 refer to Figure 8). Our premise in this article is that for medical VR training we need to stay up to L2 in order for trainees not to get distracted from the UV and focus on the learning objectives and learning outcomes instead. In the constantly evolving medical VR training landscape, there are no justifiable budgets in funds and time to overcome L3 and move towards L4 and at the same time focus on the pedagogical, learning objectives and outcomes.

Continuing from Mori’s \cite{Mori1970}, \cite{Mori2012} early identification of the “uncanny valley”, a substantial amount of research focused on discovering how visual realism could induce a negative response in the viewer \cite{Katsyri2015}. Many studies put emphasis on the evidence that the uncanny valley is a result of the mismatch in realism between elements of character design. \cite{Tinwell2014} investigated this by mismatching realism of texture and model geometry, while \cite{Zibrek2018} focused on mismatching the stimuli from separate categories (human with non-human faces). Motion, or the mismatch between realism and appearance and motion, is typically associated with the uncanny effect, since biological motion is a very strong cue by itself, and another evidence showed motion is more inaccurately perceived when the realism of the model increases \cite{Zibrek2018}. Certain areas, such as the face, are particularly vulnerable for the perceived unpleasantness when moving, as studies of \cite{Tinwell2014} showed. Other studies analyzed the effect of visual realism further by separating the realism of shape and material. Material was found to be the main predictor of appeal (e.g., blemish-free skin is most appealing) and shape the dominant predictor of realism (e.g., exaggerated features of the character are common in cartoons, but not typical for real humans). These studies detailed in \cite{Zibrek2018} show that while increasing visual realism could expose the character to a harsher judgement, realism itself can be created in different ways and is by itself not necessarily a predictor of affinity towards the character.

\begin{figure}
    \centering
    \includegraphics[width=0.75\linewidth]{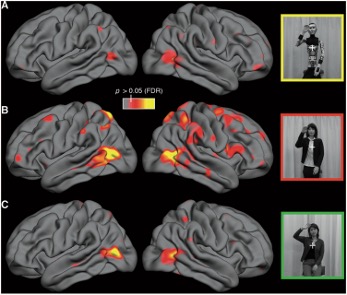}
    \caption{fMRI brain responses confirming prediction movement errors in case (B) of the robotic android as a non-human (A) or a real video (C) \cite{Saygin2012}}
    \label{fig:fig9}
\end{figure}

However, there is also a critical approach to UV from \cite{Hanson2005}, \cite{Gee2005}, \cite{Zibrek2018} that for realistic robots or virtual characters to be appealing to people, they must attain some level of integrated social responsivity, simulated personality and aesthetic refinement. An integration framework of AI, mechanical engineering and art is still needed to achieve this objective and there are several 3D design guidelines that could be tested depending on the simulation animation or 3D gaming context \cite{Tinwell2014} but have not yet been applied extensively under virtual reality that is an entirely different storytelling, “story-living” medium.

\section{Other manifestations of the Uncanny Valley (UV)}
\subsection{Consumer robots and UV}

Consumer robots are predicted to be employed in a variety of customer-facing situations. As these robots are designed to look and behave like humans, consumers attribute human traits to them—a phenomenon known as the “Eliza Effect.” In four experiments, \cite{Kim2019} show that the anthropomorphism of a consumer robot increases psychological warmth but decreases attitudes, due to uncanniness. Competence judgments are much less affected and not subject to a decrease in attitudes. That research contributes to research on artificial intelligence, anthropomorphism, and the uncanny valley phenomenon. The authors suggest to managers that they need to make sure that the appearances and behaviors of robots are not too human-like to avoid negative attitudes toward robots. Moreover, they conclude that managers and researchers should collaborate to determine the optimal level of anthropomorphism.

\subsection{UV in Haptics}

Recently the phenomenon of Uncanny Valley has been proved to also exist in haptics simulation \cite{Berger2018}. The subjective perception of haptic sensations by a human operator critically depends on the fusion of haptic and visual stimuli as a unitary percept in the human brain. If the fidelity of the haptic sensation increases but is not rendered in concordance with other sensory feedback (such as visual and auditory cues), the subjective impression of realism actually gets worse, not better, thus falling directly inside the UV (figure below).

\begin{figure}
    \centering
    \includegraphics[width=0.75\linewidth]{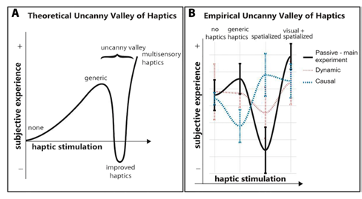}
    \caption{Uncanny Valley of haptics \cite{Berger2018}}
    \label{fig:fig10}
\end{figure}

\subsection{Uncanny Valley of interactivity for VR training}
The phenomenon of UV has also been proved to exist in medical VR training simulations too \cite{Zikas2020}. After experimenting with various design patterns and interaction techniques for VR, an interesting pattern appeared regarding the correlation of user experience and the interactivity of the VR application (Figure below). 

\begin{figure}
    \centering
    \includegraphics[width=0.75\linewidth]{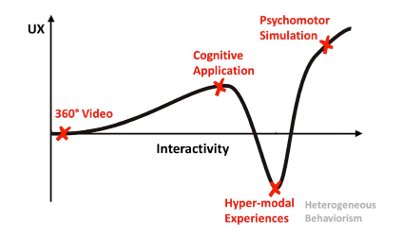}
    \caption{The Uncanny Valley of interactivity by \cite{Zikas2020}}
    \label{fig:fig11}
\end{figure}

An immersive experience relies significantly on the implemented interactive capabilities that form the user experience. As a result, to make an application more attractive in means of UX a more advanced interactive system is needed. However, as we implement more complex interaction mechanics there is a point in timeline where the UX drops dramatically. At this point, the application is too advanced and complex for the user to understand and perform the tasks with ease. We characterize this feature as heterogeneous behaviorism meaning that user’s actions do not follow a deterministic pattern resulting in the inability to complete the implemented Actions due to their incomprehensible complexity.

\section{Uncanny valley examples in Computer Animation and VR}

\subsection{Computer animation movie characters}

\begin{figure}
    \centering
    \includegraphics[width=0.65\linewidth]{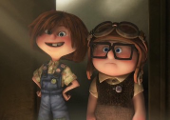}
    \includegraphics[width=0.65\linewidth]{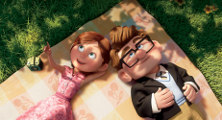}
    \includegraphics[width=0.65\linewidth]{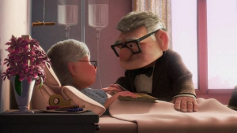}
    \caption{The computer animation movie "Up" from Pixar and the “married-life” 4-minute master-piece separate story within the main storyline, told explicitly with non-anthropomorphic, non-realistic characters and no dialogues}
    \label{fig:fig12}
\end{figure}

\textbf{Pixar’s ``Up''} immediately shattered the stereotype of computer animated movies being strictly for children. Amongst other qualities, Up’s unanimous acclaim lie in the first minutes of the film, showing young Carl Fredricksen’s fascination with exploration and his meeting a girl, Ellie, leading into a four-minute scene presenting a montage of the pair’s shared life together, famously named “Married Life https://youtu.be/9yjAFMNkCDo. The most noticeable quality of this sequence is its ability to tell a self-contained story that spans nearly 60 years without ever using one line of dialogue or anthropomorphic, or realistic CG characters. It utilizes music and visuals to tell its narrative, almost feeling like a short film in the silent era of animation, but with much more poetry than cartoons of that time would ever were known to have \cite{Taylor2014}. 

The Sony Imageworks \textbf{``The polar express''} computer animation movie is another landmark where this time fully realistic for the time CG characters were employed with full motion capture techniques for body animations and facial expressions. “Despite this characters’ friendly smiles, which appeared on billboard advertisements and a humorous film narrative, attempts to establish a rapport with the audience were lost as soon as most characters moved onscreen. E.g. The Conductor’s motion was described as puppet-like, and the audience was critical of a lack of human-likeness in his facial expression that did not match the emotive qualities of his speech. Most of the characters’ expressions also appeared out of context with a given situation as they often presented an angry expression and a cold personality when interacting with other children characters in the film” \cite{Tinwell2014}.

\begin{figure}
    \centering
    \includegraphics[width=0.85\linewidth]{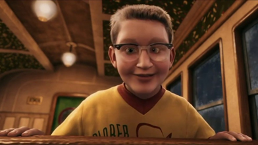}
    \includegraphics[width=0.85\linewidth]{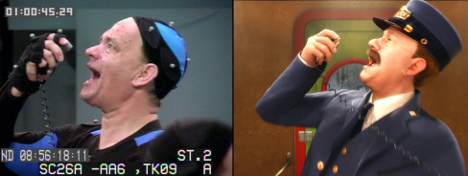}
    \caption{``The Polar Express'' computer animation movie with the most realistic up to date virtual actors, that fell right into the uncanny valley}
    \label{fig:fig13}
\end{figure}

\subsection{Networked Collaborative VR world characters}

Massive, collaborative, networked virtual environments is the next final frontier in terms of how we work, play, communicate. The two largest providers of these “metaverses” are Facebook and HTC Vive, both choosing Pixar-like non-anthropomorphic virtual characters to depict user avatars and agents in their online worlds. Facebook has been very vocal about Facebook Horizon {https://www.oculus.com/facebook-horizon/}  and disclosing that they see it as the next evolution of their social media platform. Currently in beta-form, it is striking that they fully avoid UV and stay at L2 point by employing non-realistic virtual characters, which seem that soon will define the working standard and the large population will develop a certain affinity towards this particular outlook.

HTC Vive is also following a similar approach with their Vive Sync platform, as you can see in the figures below (\footnote{\url{https://blog.vive.com/us/2020/04/30/htc-vive-opens-free-beta-vr-collaboration-app-business-vive-sync/}}):

\begin{figure}
    \centering
    \includegraphics[width=0.8\linewidth]{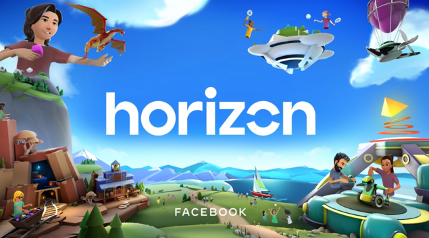}
    \includegraphics[width=0.8\linewidth]{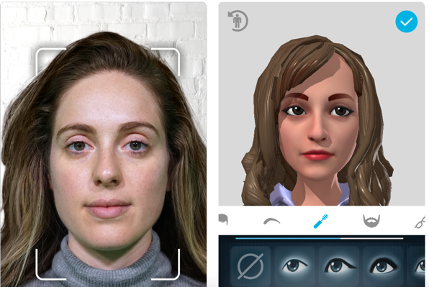}
    \caption{Facebook Horizon (top) and Vive Sync (below) that employ non-anthropomorphic/realistic virtual characters in order to avoid the uncanny valley}
    \label{fig:fig14}
\end{figure}

\section{Examples of interactive VR characters for medical training, potentially within the uncanny valley}

We are commencing this visual overview with one of our early medical VR training for emergency medicine and one of the first appearances in the bibliography, as a result of our work at the University of Geneva and the JUST EU-funded R\&D project, shown in the figure below.

\begin{figure}
    \centering
    \includegraphics[width=1\linewidth]{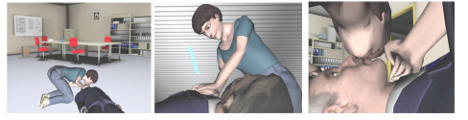}
    \caption{Emergency medicine VR training simulator featuring real-time body, motion-capture and interactive character simulation from \cite{Ponder2002}, almost 20 years ago. Someone can argue that the depicted characters lie within the UV, although it was a research prototype, pushing the state-of-the-art at that time.}
    \label{fig:fig15}
\end{figure}

In the following examples we present some of our ORamaVR medical VR training simulators involving realistic virtual humans that can be argued to fall within the Uncanny Valley:

\begin{figure}
    \centering
    \includegraphics[width=1\linewidth]{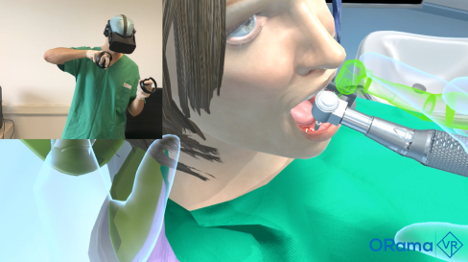}
    \caption{Dental surgical implant placement, a fully psychomotor training simulator from ORamaVR featuring a realistic, breathing virtual human with look-at, eye-contact model. It can be argued that it also falls within UV.}
    \label{fig:fig16}
\end{figure}

\begin{figure}
    \centering
    \includegraphics[width=1\linewidth]{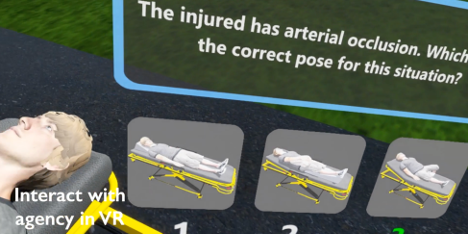}
    \caption{An emergency medicine patient transport cognitive simulator from ORamaVR, which can also be argued that falls within UV}
    \label{fig:fig17}
\end{figure}

In the following figures we present virtual human characters from \cite{1}, \cite{2} and \cite{3} that could be argued that they also fully fall within the Uncanny Valley:

\begin{figure}
    \centering
    \includegraphics[width=0.45\linewidth, height = 120pt]{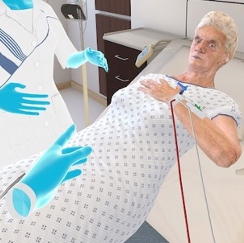}
    \includegraphics[width=0.45\linewidth, height = 120pt]{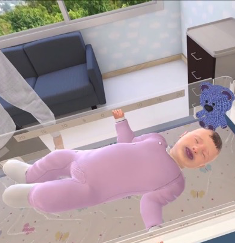}
    \caption{Medical VR cognitive training simulator. The patient character can be argued that it falls within the UV.}
    \label{fig:fig18}
\end{figure}

\begin{figure}
    \centering
    \includegraphics[width=0.45\linewidth, height = 110pt]{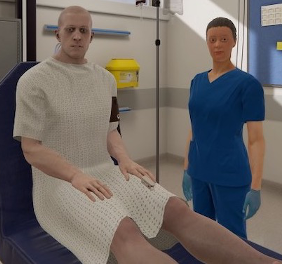}
    \includegraphics[width=0.45\linewidth, height = 110pt]{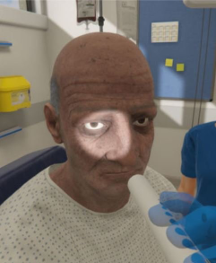}
    \caption{Medical VR cognitive training simulator.  Both patient and nurse characters can be argued that fall within the UV}
    \label{fig:fig19}
\end{figure}

\begin{figure}
    \centering
    \includegraphics[width=0.45\linewidth, height = 100pt]{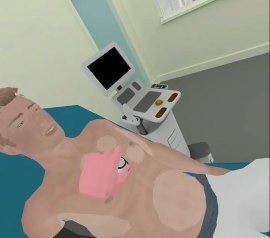}
    \includegraphics[width=0.45\linewidth, height = 100pt]{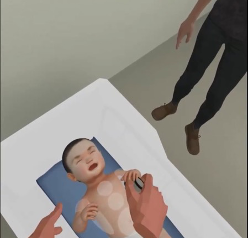}
    \caption{Medical VR cognitive training simulator.  The patient models can be argued that they fall within the UV.}
    \label{fig:fig20}
\end{figure}

\begin{figure}
    \centering
    \includegraphics[width=0.45\linewidth, height = 100pt]{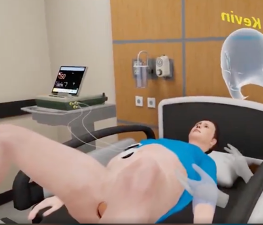}
    \includegraphics[width=0.45\linewidth, height = 100pt]{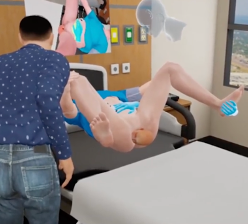}
    \caption{Medical VR birth training simulator.  The patient models can be argued that they fall within the UV}
    \label{fig:fig21}
\end{figure}

\section{The Impact of VR Hardware Limitations on the Uncanny Valley: A Case Study of Assassin’s Creed}

The uncanny valley effect, a phenomenon where virtual characters appear eerily lifelike but not quite human, causing discomfort, is becoming increasingly pronounced in virtual reality (VR) applications. This is primarily due to the limitations of current VR headsets, which include lower resolution, limited processing power, and reduced graphical capabilities compared to desktop platforms. These hardware constraints make imperfections in character realism more noticeable, leading to a more significant uncanny valley effect. Even high-budget games from major studios struggle to achieve the same level of photorealism in VR as they do on desktop platforms. To illustrate this issue, we examine the graphical evolution and challenges faced by the Assassin’s Creed series as it transitioned from desktop to VR.

\textbf{Graphical Evolution in Assassin’s Creed Series}

The budget for Assassin’s Creed games has varied significantly over the years, reflecting the increasing complexity and graphical fidelity:
\begin{itemize}
\item Assassin’s Creed One (2007): \$20 million
\item Assassin’s Creed 2 (2009): \$25 million
\item Assassin’s Creed Brotherhood (2010): \$25-30 million
\item Assassin’s Creed Rogue (2014): \$30-35 million
\item Assassin’s Creed Unity (2014): \$120 million
\item Assassin’s Creed Origins (2017): \$140-150 million
\item Assassin’s Creed Valhalla (2020): \$160-175 million
\end{itemize}

These budgets highlight the investment in achieving high graphical fidelity, particularly in creating realistic virtual humans.

\textbf{Desktop vs. VR Graphics in Assassin’s Creed}

\begin{enumerate}
\item \textbf{Desktop Graphics:}
    \begin{itemize}
        \item \textbf{Assassin’s Creed Valhalla (2020):} Building on the advancements of previous titles, Valhalla offers even more detailed character models, realistic facial animations, and dynamic lighting. The desktop platform allows for high-resolution textures and complex shaders that enhance the realism of virtual humans. The game’s characters move with lifelike fluidity and exhibit a wide range of emotions, making interactions feel more genuine and engaging.
    \end{itemize}
\item \textbf{VR Graphics:}
    \begin{itemize}
        \item \textbf{Assassin’s Creed Nexus VR:} Despite a production span of three years and a significant budget, the VR version of Assassin’s Creed faces inherent limitations in achieving photorealistic characters. The current generation of VR headsets has constraints related to processing power, display resolution, and graphical capabilities, which affect the realism of virtual humans. These limitations make the uncanny valley effect more pronounced in VR, where imperfections in character design and movement are more noticeable.
    \end{itemize}
\end{enumerate}

\textbf{Comparing Desktop and VR Versions}

\textbf{Resolution and Textures:}

\begin{itemize}
\item \textbf{Desktop:} Assassin’s Creed Valhalla benefits from high-resolution textures and detailed environments that contribute to the overall realism. Textures are rich and varied, allowing for intricate details in clothing, skin, and environmental elements.
\item \textbf{VR:} In Assassin’s Creed Nexus VR, the resolution is significantly lower to maintain performance, resulting in less detailed textures. This reduction in texture quality can make characters and environments appear more cartoonish and less immersive.
\end{itemize}

\textbf{Lighting and Shading:}

\begin{itemize}
\item \textbf{Desktop:} Advanced lighting techniques in Valhalla, such as dynamic shadows, global illumination, and complex shaders, create a visually rich environment that enhances the realism of characters and scenes.
\item \textbf{VR:} VR headsets have to compromise on lighting complexity to ensure smooth performance. This often results in simpler shading models and less dynamic lighting, which can make the environment and characters appear flatter and less realistic.
\end{itemize}

\textbf{Animations:}

\begin{itemize}
\item \textbf{Desktop:} The desktop version of Assassin’s Creed features smooth and detailed animations, including facial expressions, body movements, and interactions with the environment. These animations help in creating believable and engaging characters.
\item \textbf{VR:} Due to hardware limitations, animations in the VR version may not be as fluid or detailed. Subtle facial expressions and intricate body movements are often simplified, which can detract from the realism and exacerbate the uncanny valley effect.
\end{itemize}

The uncanny valley effect is more pronounced in VR due to the limitations in graphical fidelity imposed by less powerful processors and graphics capabilities. As virtual humans in VR strive for realism, any imperfections in animation, texture detail, or lighting become more noticeable and can lead to discomfort or unease among users. In contrast, desktop platforms, with their superior hardware, can render higher fidelity models with smoother animations and more detailed textures, helping to bridge the gap towards photorealism.

\begin{figure}
    \centering
    \includegraphics[width=0.45\linewidth]{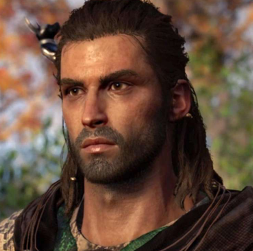}
    \includegraphics[width=0.45\linewidth]{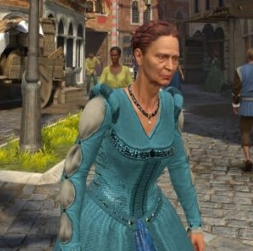}\\
    \includegraphics[width=0.45\linewidth]{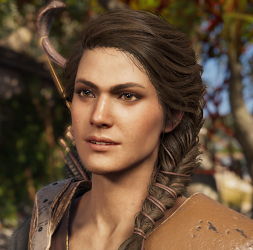}
    \includegraphics[width=0.45\linewidth]{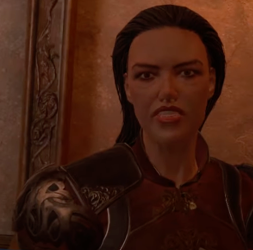}
    \caption{On the left, the desktop version (Assassin’s Creed Valhalla, 2020) showcases high graphical fidelity. On the right, the VR version (Assassin’s Creed Nexus VR) demonstrates the limitations of current VR hardware.}
    \label{fig:fig22}
\end{figure}

\section{Our proposed ``hybrid inside-out'' (HIO) rendering methodology for virtual reality characters away from the Uncanny valley}

In this section we provide a visual overview of our HIO rendering methodology at ORamaVR in order to avoid Uncanny Valley effects. This novel methodology introduces non-photorealistic modelling \& rendering of non-anthropomorphic embodied conversational agents, avatars and AI agents. With this technique, early qualitative user study evaluations have shown that trainees can focus more on the learning objectives and the training outcomes and not be distracted by the UV effects. A key insight of HIO is that non-anthropomorphic external character appearance (out) and realistic internal anatomy (inside) can be combined in a hybrid VR rendering simulation approach for networked, deformable, interactable psychomotor and cognitive training simulations.

\begin{figure}
    \centering
    \includegraphics[width=0.9\linewidth]{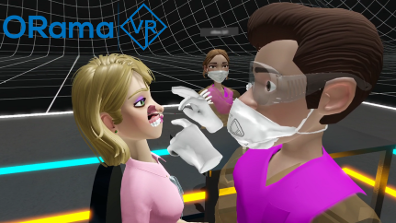}
    \includegraphics[width=0.9\linewidth]{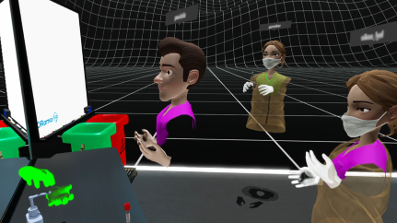}
    \caption{The ORamaVR HIO methodology applied to a Covid-19 PPE and swab testing VR training simulator \cite{CVRSB}.}
    \label{fig:fig23}
\end{figure}

\begin{figure}
    \centering
    \includegraphics[width=0.9\linewidth]{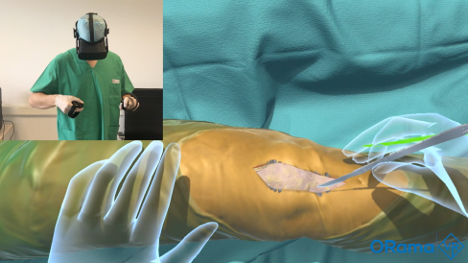}
    \includegraphics[width=0.9\linewidth]{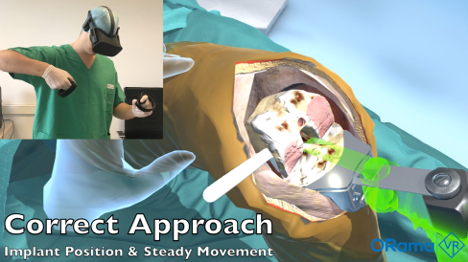}
    \caption{Physically based anatomical rendering in order for surgeons to easily and clearly identify pathology, employing the HIO methodology by ORamaVR, in surgical, fully psychomotor Total Knee and Hip Arthroplasty VR training simulators}
    \label{fig:fig24}
\end{figure}

\begin{figure}
    \centering
    \includegraphics[width=0.9\linewidth]{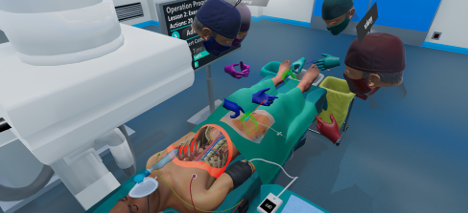}
    \includegraphics[width=0.9\linewidth]{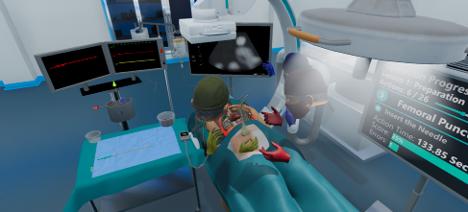}
    \caption{Hybrid non-anthropomorphic characters (outside) and physically based anatomical rendering (inside, as another complex case study of the HIO methodology from ORamaVR, applied in a REBOA emergency medicine procedure}
    \label{fig:fig25}
\end{figure}

\begin{figure}
    \centering
    \includegraphics[width=0.9\linewidth]{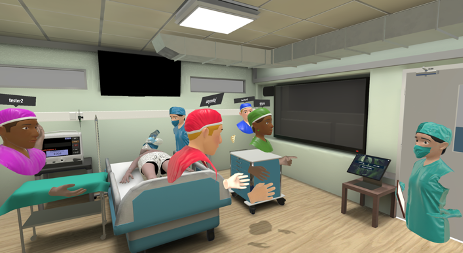}
    \includegraphics[width=0.9\linewidth]{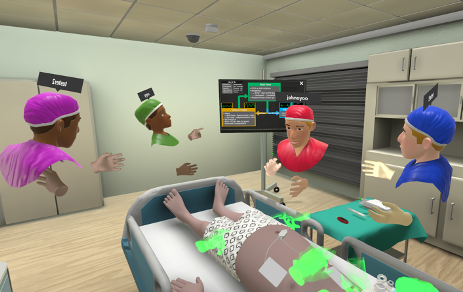}
    \caption{Multiple avatars and characters engaged in a Collaborative Cardiac Arrest Resuscitation Simulation in VR, demonstrating the application of the Hybrid Inside-Out (HIO) rendering methodology to enhance realism and reduce the Uncanny Valley effect \cite{iReact}. Created using MAGES 4.0 \cite{MAGES4}.}
    \label{fig:fig26}
\end{figure}

\begin{figure}
    \centering
    \includegraphics[width=0.9\linewidth]{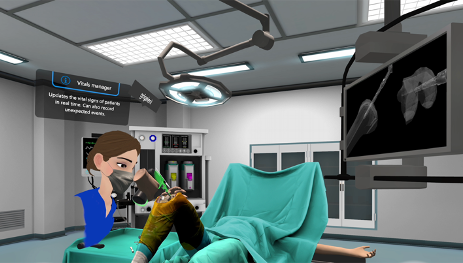}
    \includegraphics[width=0.9\linewidth]{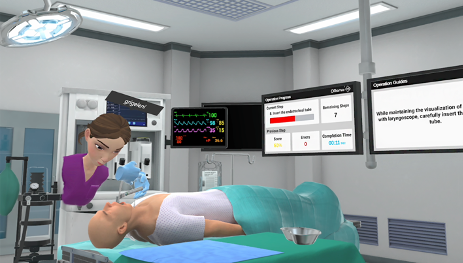}
    \caption{Virtual human characters performing medical procedures in a VR environment, showcasing the use of high-fidelity texture mapping and dynamic lighting adjustments as part of the HIO rendering methodology.}
    \label{fig:fig27}
\end{figure}

\section{Conclusions}

A number of early observations have already been drawn from all examples above. 

\textbf{Design Elements and Realism:} Design elements should match in human realism only when considered outside-in. A CG character does not appear uncanny when human anatomy inside models and non-anthropomorphic external elements are mixed. Still, matching appearance and motion kinematics are important elements.

\textbf{Reducing Conflict and Uncertainty:} Reducing conflict and uncertainty by matching appearance, behavior, and ability in the appropriate context is crucial. A virtual character’s appearance or demeanor systematically influences people’s perceptions of the character and their willingness to comply with the character’s instructions. These perceptions and responses are evidently elicited by social cues embodied in the character and are framed by people’s expectations of their role in the situation. For example, participants in user evaluation studies did not find the more humanlike, attractive, or playful robot more compelling across the board. Instead, they expected the robot to look and act appropriately, given the task context. Thus, if a CG character appears too appliance-like, people expect little from it; if it looks too human, people expect far more from it. A highly human-like appearance leads to an expectation that certain behaviors are present, such as humanlike motion dynamics.

\textbf{Open Research Fields:} Research topics in our field such as facial expressions and visemes, body postures, hand gestures, lip-sync, eye-contact, gaze models, speech synthesis, breathing simulation instead of stillness, simulated empathy, personality traits, and emotional conversation for virtual character interaction in medical VR, are still open research fields. Further research is needed to improve or potentially overcome the uncanny valley in medical VR. It is our firm belief that now more than ever this task is possible.

There is a growing lack of medical professionals globally (40 million by 2030) and not enough are trained for these future needs. Even access to quality medical services today is still very limited - 5 billion people are lacking affordable surgical or anesthesia care due to high medical costs or shortage of physicians. The problem lies in the outdated paradigm of medical training that has not changed for 150 years despite technological advances: a master operates on a patient and the trainee/apprentice observes while then learns on-the-job, practicing on real patients. This is a dangerous, costly, lengthy, and non-scalable training process for the future surgeons, doctors, and nurses whose turnover is an important problem \cite{Shader2001}.

At ORamaVR we aim to transform medical education by enabling the rapid construction of the world’s most intelligent, symbiotic Virtual Reality (VR) training simulations, at a fraction of the time and cost than anyone else in this field. Today, building a single high-quality virtual training module takes months and costs tens of thousands of Euros. This makes it a challenge to build, maintain, and continuously update it to keep up with medical evolution. We are solving the most difficult technical problems that make medical VR content creation so complex and costly. The ultimate goal of ORamaVR is to enable high-quality content that is built code-free/low-code by medical professionals themselves, using our proprietary, innovative platform that relies on breakthrough VR science. ORamaVR (www.oramavr.com) is a deep technology, spatial computing startup, innovating the field of med-ed-tech Virtual Reality (VR) training simulation platforms. Our technologies are based on state-of-the-art research by world-renowned scientists team members. This research is turned into a novel, symbiotic VR authoring and simulation platform by our top engineers, medical professionals, and business developers.
Despite these and other advances, the Uncanny Valley when interacting with virtual characters in medical VR is a real phenomenon, and it is further exacerbated by the current limitations of VR technology, including lower resolution, limited processing power, and reduced graphical capabilities. As amongst the leading organizations in medical VR training, it is our responsibility to increase awareness of this fact to all involved partners. The aim is that all can identify it clearly so that we can then collectively act upon and improve our medical VR training simulators. This is our individual as well as collective responsibility for the rapid acceleration of human learning and the next generation of medical professionals, especially in a post-pandemic era.


\section{Future Work}

Looking ahead, several areas warrant further exploration to continue improving the realism and effectiveness of VR applications, particularly in mitigating the Uncanny Valley effect. Future research should focus on developing more powerful VR hardware with enhanced processing capabilities and graphics performance to support higher fidelity textures, lighting, and animations. The development of next-generation headsets that provide higher resolution displays, faster refresh rates, and more accurate tracking systems will be crucial in bridging the gap between VR and desktop graphics quality. Further integration of machine learning algorithms to dynamically adjust and enhance character realism in real-time, adapting to user interactions and environmental conditions, will also be key. Conducting comprehensive user experience studies to better understand the factors contributing to the Uncanny Valley effect and how users perceive and interact with virtual characters is essential. These studies should focus on diverse user groups to ensure that the solutions developed are broadly applicable and effective. Expanding the comparison of VR and desktop platforms across different genres and applications will help identify specific areas for improvement in VR technology. Encouraging collaboration between experts in computer graphics, VR, and medical training can lead to the creation of more sophisticated and effective training tools. Additionally, focusing on creating scalable VR training modules that can be built and maintained cost-effectively while keeping up with medical advancements will democratize the creation of VR content, allowing more institutions to benefit from high-quality training simulations. By pursuing these avenues, we can continue to enhance the realism of virtual humans and improve the overall effectiveness and immersion of VR applications in various fields, particularly in medical education and training. The ultimate goal is to create a seamless, immersive learning experience that prepares medical professionals for real-world challenges, thereby improving patient outcomes and advancing the field of medical education.

\bibliographystyle{spmpsci}

\end{document}